\documentstyle[multicol,aps,psfig]{revtex}
\begin{document}
%\draft
\preprint{LBNL-49561}

\title{LPM Interference and Cherenkov-like Gluon Bremsstrahlung in Dense Matter}
\author{A. Majumder and Xin-Nian Wang}
\address{Nuclear Science Division, MS 70-319,
Lawrence Berkeley National Laboratory, Berkeley, CA 94720 USA}

\date{\today}

\maketitle

\vspace{-1.5in}
{\hfill LBNL-58446}
\vspace{1.4in}

\begin{abstract}
Gluon bremsstrahlung induced by multiple parton scattering in a 
finite dense medium has a unique angular distribution 
with respect to the initial parton direction. 
A dead-cone structure with an opening angle 
$\theta^2_0 \approx 2(1-z)/(zLE)$ for gluons with fractional
energy $z$ arises from the Landau-Pomeranchuck-Migdal (LPM)
interference. In a medium where the gluon's dielectric constant is
$\epsilon >1$, the LPM interference pattern is shown to become
Cherenkov-like with an increased opening angle determined by the
dielectric constant $\cos^2\theta_c=z+(1-z)/\epsilon$. 
For a large dielectric constant $\epsilon \gg 1+2/z^2LE$, the
corresponding total radiative parton energy loss is about twice that
from normal gluon bremsstrahlung. Implications of this Cherenkov-like 
gluon bremsstrahlung to the jet correlation pattern in high-energy heavy-ion 
collisions is discussed.

\noindent {\em PACS numbers:} 12.38.Mh, 24.85.+p; 13.60.-r, 25.75.-q
\end{abstract}
\pacs{ 12.38.Mh, 24.85.+p, 25.75.-q, 11:80La}

\begin{multicols}{2}

\section{Introduction}

The two most striking experimental observations in central $Au+Au$ 
collisions at the Relativistic Heavy-ion Collider (RHIC) are the
large collective flow \cite{v2}  and strong jet 
quenching \cite{phenix-r,star-jet} that are believed 
to be the experimental evidence \cite{whitepapers,gm} for 
the formation of strongly interacting quark-gluon plasma (sQGP). The 
observed strong collective flow is measured by the azimuthal anisotropy $v_2$
which results from the initial energy density or pressure and geometric
ellipticity of the dense matter \cite{ollitrault}. The measured $v_2$
was found to reach the hydrodynamic limit of a perfect fluid with extremely
small viscosity \cite{teaney,heinz}. Jet quenching or the suppression of
high $p_T$ hadron spectra, on another hand, is caused by parton
energy loss \cite{wg92} as the energetic parton jet propagates through
the dense medium. Phenomenological analyses of the jet quenching
pattern based on parton energy loss reveal an initial
parton density that is about 30-40 times higher than in a cold nuclear
matter \cite{ww02,gv02,wang03}. Such a density is also consistent with 
the hydrodynamical analysis of the observed collective flow \cite{heinz}.

In an effort to recover the energy lost by the leading hadrons, one
finds indeed an enhancement of soft hadrons along the direction
of the initial parton jets \cite{star-cone,phenix-cone} as determined by
the opposite direction of the triggered high $p_T$ hadron. These soft
hadrons, however, have a much broadened angular distribution which
peaks at a finite angle away from the initial jet direction. This is
quite different from the distribution in $pp$ collisions which peaks
along the direction of the initial jet. Such a phenomenon has
been attributed to Mark cone or conical 
flow \cite{shuryak-cone,chaudhuri,stocker} caused by the propagation 
of a supersonic jet through the dense medium. The same angular pattern
could also be a result of Cherenkov gluon radiation \cite{ruppert,dremin}.
However, as pointed out in Ref.~\cite{kmw} and will be discussed later in
this paper, the total energy loss caused by Cherenkov gluon radiation
is very small as compared to radiative energy loss induced by multiple
parton scattering \cite{gw1,zhak,bdms,glv,wied,gw01}. Therefore, it
cannot be the dominant cause of the observed jet quenching and the
induced soft hadron production.

This paper will discuss the angular distribution of gluon
bremsstrahlung induced by multiple scattering of a fast parton
in a medium that has a gluonic dielectric constant $\epsilon>1$
and the consequences on the total radiative energy loss. 
We start with an analysis of the bremsstrahlung of light-like gluons 
induced by multiple scattering that has a unique angular distribution
determined by the gluon formation time relative to the medium size
due to the Landau-Pomeranchuck-Migdal (LPM) interference. The results are
then extended to the case when gluons acquire a space-like dispersion
relation in the medium due to a large dielectric constant $\epsilon>1$.
The angular distribution of the space-like gluon bremsstrahlung
will then be shown to peak at an angle solely determined by the 
gluon dielectric constant $\epsilon$ in the medium. Such a 
Cherenkov-like pattern is shown to be the result of the LPM 
interference in induced bremsstrahlung
of gluons with a large dielectric constant or index of 
refraction $n=\sqrt{\epsilon}$. Other features of the Cherenkov-like 
gluon bremsstrahlung and consequences on the soft particle distribution 
induced by jet quenching in high-energy heavy-ion collisions
will also be discussed.

\section{LPM Interference and the dead-cone}

Gluon bremsstrahlung induced by multiple scattering of a fast
parton in a dense medium has been shown to possess many interesting
features \cite{gw1,zhak,bdms,glv,wied,gw01} due to the LPM interference 
in QCD. One of the unique features is the quadratic length ($L$) 
dependence of the parton radiative energy
loss in a finite and static medium. This is a
consequence of the LPM interference and the unique feature of
gluon radiation in QCD. The most important
contributing factor is the interaction of the gluonic cloud surrounding
the propagating parton with the medium and the resulting gluon
bremsstrahlung. Such non-Abelian LPM interference should also be
manifested in the underlying final gluon spectra.

In the framework of twist expansion, the gluon spectra induced by
double scattering of a fast quark with energy $E$ in a static dense 
medium has been obtained in Ref. \cite{gw01} as,
\begin{equation}
\frac{dN_g}{dzd\ell_T^2}=P(z)
\alpha_s^2\widetilde{C}m_NL
\frac{1}{\ell_T^4}\left[ 1-e^{-(L/\tau_f)^2}\right],
\label{eq1}
\end{equation}
where $z$ and $\ell_T$ is the gluon's fractional energy
and transverse momentum, respectively. $P(z)=[1+(1-z)^2]/z$ is
the quark-gluon splitting function and $\tau_f=2z(1-z)E/\ell_T^2$
is defined as the gluon's formation time. The above result is obtained 
for a quark propagating through a cold nucleus with a Gaussian density
distribution $\rho(r)\sim \exp(-r^2/2L^2)$. The parameter $\widetilde{C}m_N$ 
represents the gluon correlation strength and $m_N$ is the nucleon 
mass introduced only for the convenience of normalization.
In a hot QCD medium, $\alpha_s\widetilde{C} m_N \sim \mu^2\sigma_g\rho_0$
\cite{ww02} with $\mu$ the Debye screening mass, $\sigma_g$ the
parton-gluon cross section and $\rho_0$ the average gluon density.
Here, one has used the relation $\alpha_s xG(x) \sim \mu^2\sigma_g$
as the gluon density probed by the propagating parton in the medium.

\begin{figure}
\centerline{\psfig{figure=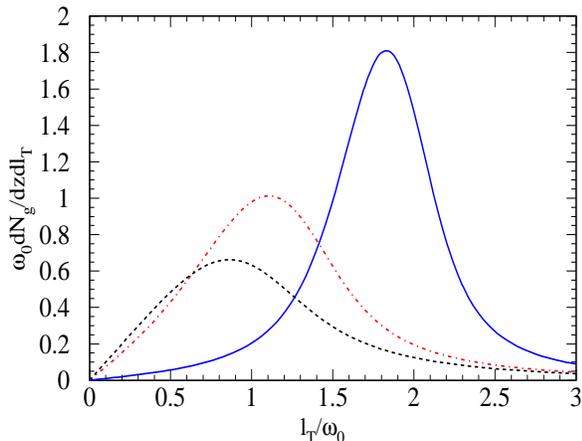,width=3.0in,height=2.3in}}
\caption{Angular distribution of gluon bremsstrahlung induced by
multiple parton scattering: $\omega_0dN_g/dzd\ell_T$ versus $\ell_T/\omega_0$,
where $\omega_0\equiv \sqrt{2Ez(1-z)/L}$. Gluon spectra is
scaled by $P(z)\widetilde{C}\alpha_s^2Lm_N/\omega_0^2$ and the gluon's
fractional momentum is set $z=0.2$. The dashed line is
for ordinary gluon bremsstrahlung with LPM interference as given
by Eq.~(\ref{eq1}). The solid and dot-dashed lines are the distribution 
in Eq.~(\ref{chrk}) for the bremsstrahlung of gluons that has a space-like 
dispersion relation with $-\Pi_T(\ell)/\omega_0^2=4$ and 1, respectively.}
\label{fig1}
\end{figure}

The above spectra is finite at $\ell_T=0$ in contrast to the
gluon bremsstrahlung in vacuum which is collinearly divergent. This
is a unique feature of LPM interference in a non-Abelian theory.
Because of the LPM interference, gluons with formation time
$\tau_f=2z(1-z)E/\ell_T^2$ much larger than the medium size $L$
will have destructive interference and will be suppressed.
This will lead to a depletion of gluons in the forward cone
within an angle 
\begin{equation}
\theta_{0}\approx \frac{\omega_0}{zE}\equiv\sqrt{\frac{2(1-z)}{zEL}},
\end{equation}
as illustrated in Fig.~1 by the dashed line. Therefore, LPM
interference effectively creates a dead-cone for the gluon
bremsstrahlung in the direction of the fast parton, as also
pointed in Ref.~\cite{vitev}. The size
of the dead-cone decreases both with the parton energy and
medium size. The angular distribution of the bremsstrahlung
gluons is peaked at $\theta_{0}$ and the width of the peak is
also about $\theta_{0}$.

In practice, the collinear divergence in parton fragmentation will
be regulated by the hadronization scale $\ell_T \sim \Lambda_{\rm QCD}$.
This means any angular structure within an 
angle $\theta<\Lambda_{\rm QCD}/zE$ cannot be observed. This requires
$zE/L> \Lambda_{\rm QCD}^2/2$ in order for the LPM interference
pattern to be observed in the soft hadron spectra.

\section{Cherenkov-like gluon radiation}

For a more complete treatment of the gluon bremsstrahlung in
hot QCD medium, one should also include the effect of further
interaction between the radiated gluon and the medium. At high
temperature in QCD, for example, one can simply replace the
gluon propagator by an effective one as one resums all hard
thermal loops \cite{htl} as has been discussed in Ref.~\cite{mag}
in the calculation of radiative energy loss by a heavy quark.
If the temperature of the QGP is just above $T_c$, hard
thermal loop (HTL) treatment of parton interaction at finite
temperature is known to fail to describe lattice QCD 
results \cite{blaizot,strickland}.
One might expect a different dispersion relation for gluons
in this region where the strong interaction between partons are
non-perturbative. The empirical observation of the sQGP also indicates
the non-perturbative behavior of QCD matter as created in the central
$Au+Au$ collisions at RHIC. These observations have led to the suggestion 
that the QCD matter could become effectively composed of medium-modified
(heavy) quarks and gluons and their screened Coulomb potential could 
lead to many shallowly bound states \cite{shuryak-zahed}. 
The strong interaction between these bound states can provide an effective
mechanism for the observed small viscosity. These bound states most 
likely exist in the gluon sector or could be quite heavy, since abundant 
light quark bound states can be ruled
out by the lattice study of charge and baryon number 
fluctuation  or strangeness-baryon correlations \cite{majumder}.
If such gluonic or heavy bound states exist, the interaction
between gluon and these bound states could lead to a space-like
dispersion relation that gives rise to a gluon dielectric constant
$\epsilon >1$. A recent study within a simple model of transitional
excitation of heavy particles by a light particle indeed shows
a space-like dispersion relation for the light particle in the soft
region \cite{kmw}. Such an effective gluonic dispersion relation
was also suggested in a recent study \cite{ruppert} of Mach-cone-like 
density excitation by a space-like longitudinal plasmon mode 
(or Cherenkov radiation).

For the purpose of studying the pattern of Cherenkov-like gluon radiation
induced by multiple scattering in this paper, let us simply assume an 
effective space-like dispersion relation for gluons. Therefore, the gluon 
propagator will have a general form
\begin{equation}
D^{\mu\nu}(\ell)=-\frac{{\cal P}_T^{\mu\nu}}{\ell^2-\Pi_T+i\epsilon}
-\frac{{\cal P}_L^{\mu\nu}}{\ell^2-\Pi_L+i\epsilon},
\end{equation}
where ${\cal P}_T^{\mu\nu}$ and ${\cal P}_L^{\mu\nu}$ are the transverse
and longitudinal projector, respectively. As an illustration of
the consequences of the space-like gluon dispersion relation in this
paper, we focus only on the transverse part in the calculation of
induced gluon bremsstrahlung. We will also use the corresponding spectral
function for the final gluons. We assume ${\rm Re}\Pi_T(\ell)<0$ 
in the dispersion relation $\ell^2-\Pi_{T}(\ell)=0$ and 
that the imaginary part 
${\rm Im} \Pi_T(\ell) \ll 2[\vec{\ell}^2+ {\rm Re}\Pi_T(\ell)]$ \cite{kmw} 
in the regime of our interest so that these soft gluons are not damped 
during the propagation through the medium. With these simplifications, the 
calculation of induced gluon bremsstrahlung via multiple scattering 
is very similar to the normal case. The final gluon distribution can 
be obtained from Eq.~(\ref{eq1}) with the replacement
$\ell_T^2\rightarrow \ell_T^2+(1-z)\Pi_T(\ell)$,
\begin{equation}
\frac{dN_g}{dzd\ell_T^2}\approx P(z)
\frac{\alpha_s^2\widetilde{C}L}{m_N}
\frac{1}{[\ell_T^2+(1-z)\Pi_T(\ell)]^2}
\left[ 1-e^{-(L/\widetilde{\tau}_f)^2}\right],
\label{chrk}
\end{equation}
where $\widetilde{\tau}_f=2Ez(1-z)/[\ell_T^2+(1-z)\Pi_T(\ell)]$.

The above gluon spectra clearly has a very different structure
for a space-like dispersion relation $\Pi_T(\ell)<0$ from
that in Eq.~(\ref{eq1}). It is peaked at $\ell_T^2=(1-z)|\Pi_T(\ell)|$
which is independent of the medium size $L$, in contrast to the
angular distribution of bremsstrahlung of normal light-like gluon with 
LPM interference. The corresponding angular distribution has also
a peak structure and is strongly suppressed in the forward 
direction within a cone $\theta<\theta_c$ as illustrated by
the solid and dot-dashed lines in Fig.~1. The width of the peak is, 
however, the same as the bremsstrahlung of ordinary light-like gluons 
with LPM interference. The shift of the peak to a larger angle
depends on the value of $\Pi_T(\ell)/\omega_0^2$.
For simplicity, we have set $\Pi_T(\ell)$ as independent of $\ell_T$
for soft gluons in Fig.~1 as an illustration. In general $|\Pi_T(\ell)|$
should depend on gluon's momentum $\ell$ and decreases with $\ell$
at large momentum. In this limit, gluons will become light-like again.
With a momentum-dependent $|\Pi_T(\ell)|$, the shape and position of 
the peak will be slightly modified.

The above pattern of gluon bremsstrahlung is very similar to that of 
Cherenkov radiation. However, when cast in the form of induced radiation
from multiple parton scattering, one can immediately discover a special
relationship between the Cherenkov-like gluon radiation pattern
and the LPM interference effect. In our case of induced gluon radiation, 
the Cherenkov-like bremsstrahlung is a result
of complete LPM interference within the forward cone $\theta_c$.
If we express the self-energy of a space-like
gluon in terms of gluon's dielectric constant $\epsilon(\ell)$,
\begin{equation}
\epsilon(\ell)\equiv 1-\frac{\Pi_T(\ell)}{\ell_0^2},
\end{equation}
or $\epsilon(\ell)=\vec{\ell}^2/\ell_0^2$, one can find the cone-size of 
this Cherenkov-like gluon radiation as
\begin{equation}
\cos^2\theta_c=z+\frac{1-z}{\epsilon(\ell)}.
\end{equation}
In the soft radiation limit $z\sim 0$, this corresponds exactly
to the angle of classical Cherenkov radiation.

One can also compute the total quark energy loss due to the Cherenkov-like
gluon radiation from Eq.~(\ref{chrk}),
\begin{eqnarray}
\Delta E&=&E \int dz d\ell_T^2 z\frac{dN_g}{dzd\ell_T^2} \nonumber \\
&\approx &\Delta E_0 \times \left\{ \begin{array}{ll}
    2, & {\rm for} \frac{|\Pi_T|L}{2E} \gg 1 \\
    \left(1+\frac{1}{3}\frac{|\Pi_T|L}{2E}\right), &  
    {\rm for} \frac{|\Pi_T(\ell)|L}{2E} \ll 1
    \end{array} \right.
\end{eqnarray}
where $\Delta E_0\approx \widetilde{C}\alpha_s^2m_N L^2 3\ln(E/\mu)$ is
the radiative energy loss from normal gluon bremsstrahlung \cite{ww02} 
with gluon spectra as given by Eq.~(\ref{eq1}) and $\mu$ is the averaged 
transverse momentum transfer for elastic parton scattering in the
medium. For simplification, we assumed that $\Pi_T(\ell_0)\approx \Pi_T$
has a weak momentum dependence for soft (space-like) gluons. 
It is interesting to note
that the total energy loss becomes twice of that from normal gluon
radiation when $|\Pi_T(\ell_0)|\gg 2E/L$, which corresponds
to a large gluon dielectric constant $\epsilon \gg 1+2/z^2EL$.
In this case, the dielectric property of the medium amplifies the
induced radiative energy loss.

We note, as pointed out in Ref.~\cite{kmw}, that true Cherenkov gluon
radiation without multiple parton scattering can also cause radiative
energy loss. Its value, $(dE/dx)_c \sim 4\pi\alpha_s \ell_0^2/2$ is
considerably smaller than the scattering-induced radiative energy loss
since the typical space-like gluon energy $\ell_0 \sim T$ is rather 
soft, on the order of the temperature, when Cherenkov radiation is 
the strongest. It also has a very weak dependence on the matter density.

\section{Discussions}

We should emphasize that the gluon spectra in this paper is obtained
through a simplified treatment of induced radiation of gluons
with dielectric constant $\epsilon>1$. A more complete study
is needed, however, including the longitudinally polarized gluons, for more
accurate calculation of the final gluon spectra. However, we expect
the main feature of our study in this paper will remain. A large gluon
dielectric constant will lead to a Cherenkov-like gluon bremsstrahlung
due to LPM interference. If such a medium with large gluon dielectric
constant is created in high-energy heavy-ion collisions, then one
should see such an angular distribution of soft hadrons
in the direction of quenched jets. Such an angular distribution
appears similar to that from a Mach cone caused by a supersonic
jet. However, the underlying mechanisms are completely different and
the distribution of produced particles is determined by 
different properties of the medium. 
The pattern of Cherenkov-like gluon bremsstrahlung is determined
by the gluon dielectric constant in the medium while the cone
size of the sonic shock wave is directly related to the sound velocity in
the medium.

The momentum dependence of the cone size is also different between
sonic shock wave and the Cherenkov-like bremsstrahlung. Since the sonic
shock wave is caused by the propagation of  energy at the sound
velocity in the medium, the resulting Mach cone size should be
independent of the final soft hadrons' momenta. On the other hand,
the gluon dielectric constant has a strong dependence on the gluon
momentum. In general, it decreases with the gluon momentum, so that large
momentum gluons will become light-like again. Therefore, one expects
to see the Cherenkov cone size become smaller and the cone will
eventually disappear for high-energy gluons. 
This means that the angular correlation of
soft hadrons will gradually become peaked in the direction of
the jet when the momentum of the soft hadrons is increased.
Such a difference in the  momentum dependence of the angular cone size
can be used to distinguish the sonic shock wave from Cherenkov-like
gluon bremsstrahlung. Confirmation and measurement of such
Cherenkov pattern will provide information on the fundamental
and intrinsic properties of the dense medium.

As has been demonstrated, the Cherenkov-like gluon 
bremsstrahlung will increase
the total parton energy loss by almost a factor of two if the
gluon dielectric constant is really large. This will certainly
affect the modification of the leading hadron spectra from
fragmentation of the leading partons. One therefore has to
revise the previous analyses \cite{ww02,gv02,wang03} of the RHIC 
data on high-$p_T$ hadron suppression and the extraction of the initial 
gluon density.

We would like to thank V. Koch for helpful discussions. This work is 
supported  by the Director, Office of Energy
Research, Office of High Energy and Nuclear Physics, Divisions of 
Nuclear Physics, of the U.S. Department of Energy under Contract No.
DE-AC02-05CH11231.

%\begin{multicols}{2}

\end{multicols}

\end{document}